\documentclass[prd]{revtex4}
\usepackage{graphicx}

\def\be{\begin{equation}}
\def\ee{\end{equation}}
\def\bea{\begin{eqnarray}}
\def\eea{\end{eqnarray}}

\def\dt{\partial_\tau}
\def\dz{\partial_\eta}
\def\dtt{\partial_{\tau\tau}}
\def\dzz{\partial_{\eta\eta}}
\begin{document}
\title{Field dynamics and kink-antikink production in rapidly expanding systems}
\author{G. Holzwarth\footnote{Permanent address:
Fachbereich Physik, Universit\"{a}t Siegen, 
D-57068 Siegen, Germany\\
e-mail: holzwarth@physik.uni-siegen.de}}
\address{Institute of Theoretical Physics, University of Stellenbosch, 7602 Matieland, South Africa} 
\begin{abstract}\noindent
Field dynamics in a rapidly expanding system is investigated by transforming 
from space-time to the rapidity - proper-time frame. The proper-time dependence of different
contributions to the total energy is established. For systems characterized by a 
finite momentum cut-off, a freeze-out time can be defined after which the
field propagation in rapidity space ends and the system decays into decoupled solitons,
antisolitons and local vacuum fluctuations. Numerical simulations of field evolutions
on a lattice for the (1+1)-dimensional $\Phi^4$ model illustrate the general results and show that 
the freeze-out time and average multiplicities of kinks (plus antikinks) produced in 
this 'phase transition' can be obtained from simple averages over the initial ensemble of
field configurations. An extension to explicitly include additional dissipation is discussed.
The validity of an adiabatic approximation for the case of an overdamped system is investigated.
The (3+1)-dimensional generalization may serve as model for baryon-antibaryon production after heavy-ion
collisions. 
\end{abstract} 
\maketitle
\vspace{3cm}
\leftline{PACS numbers: 11.10.Lm,11.27.+d,25.75.-q,64.60.Cn,75.40.Mg}  
\leftline{Keywords: Rapid expansion, topological textures, heavy-ion collisions, freeze-out}

\newpage

\section{Introduction}
It is a commonly accepted scenario for central ultra-relativistic heavy-ion collisions 
that immediately after the collision, sandwiched between the receding baryonic pancakes,
a rapidly expanding spatial region is left, characterized by high energy density
and low baryon density \cite{Bjork}. The expansion may be mainly in beam direction,
such that the longitudinal velocities of volume elements are proportional to their 
distance from the collision point. Assuming invariance with respect to transformations
between comoving frames, the rapidity density of particles emitted from this expanding rod will
show a plateau reflecting that symmetry. This plateau is experimentally well established, 
and recent results for baryonic yields indicate that for high centrality and increasing energy the 
antibaryon/baryon ratio in fact tends towards unity \cite{Star,Brahms1,Brahms2}.  
 
The production of baryon-antibaryon pairs from regions of highly excited 'vacuum' presents a challenge
for many different types of theoretical models. It arises very naturally in models where
baryon density is identified with topological winding density in effective meson field theories 
\cite{DeGrand,EllKo,EllKar,KapWong}.
For random field configurations on random triangular lattices the probability for nontrivial
winding provides an upper (Kibble) limit for baryon-antibaryon multiplicities \cite{Kibble,Christ}. 
The dynamical
evolution in time, which follows the preparation of a statistical ensemble at some initial time,
tends to smoothen the random initial configurations, resulting in frequent 'annihilation' of
positive and negative winding density, such that the final baryon-antibaryon multiplicities will
strongly depend on the cooling mechanism and freeze-out times, after which the system is sufficiently
decoupled, such that baryons and antibaryons no longer interact \cite{Holzwarth,HolKlom,Zapo,RutZakZap}.

Apart from genuine energy dissipation through meson emission, the rapid expansion provides a
kinetic cooling mechanism 
which naturally decreases all gradients in a field configuration and,
consequently, the winding densities. This expansion is conveniently accounted for by transforming
the space-time frame into the rapidity - proper-time frame and formulating the field dynamics
in these locally comoving frames \cite{Huang,Cooper,Lampert,Randrup96,Randrup97,Petersen,Scavenius}. 
We repeat the essential points of this formulation in section II,
with specific attention to the soliton solutions. We also include the effects of genuine damping 
and discuss the adiabatic approximation which applies for overdamped systems and provides 
significant simplifications in the numerical simulation of such evolutions. 

In section III of the present paper we apply this formalism to the (1+1) dimensional $\Phi^4$ model,
which is characterized by only two discrete degenerate vacua. This provides a very transparent
example for the mechanisms which govern the 'phase transition' from the hot initial configurations
through the freeze-out into the final cold system which has settled into the true vacua
and consists of non-interacting solitons and vacuum fluctuations. We obtain simple analytical expressions 
for freeze-out time and kink multiplicities, both for the system moving freely in the expanding Bjorken rod, 
and for the damped system which additionally looses energy through a dissipative term.    

\section{Field dynamics in the rapidity - proper-time frame}
\subsection{Transformation to Bjorken frame}
\noindent
As standard effective action $\cal S$ for a scalar field $\Phi$ in space-time $(z,t)$ we consider
\be \label{1}
{\cal S}=\int
\left(\frac{1}{2}(\partial_t\Phi)^2-\frac{1}{2}(\partial_z\Phi)^2 
-V(\Phi)\right) dzdt
\ee
where $V(\Phi)$ denotes a suitable local self-interaction.
The transformation from $(z,t)$ to proper time $\tau$ and rapidity $\eta$ is defined as
\begin{eqnarray}
t&=&\tau \cosh\eta,~~~~~~~~~~\tau~=~\sqrt{t^2-z^2},~~~~~~~~~~~~~
\partial_t~=~\cosh\eta\; \dt -\frac{\sinh\eta}{\tau}\dz \nonumber\\
z&=&\tau \sinh\eta,~~~~~~~~~~\eta~=~\mbox{atanh}\left(\frac{z}{t}\right),~~~~~~~~~~~~
\partial_z~=~-\sinh\eta\; \dt +\frac{\cosh\eta}{\tau}\dz.
\end{eqnarray}
The boundaries in the forward light cone are (note: $dzdt=\tau d\tau d\eta$)
\be
\int_0^\Lambda dt \int_{-t}^t dz \Longrightarrow \int_0^\Lambda \tau d\tau \int_
{-\rm atanh\sqrt{1-(\tau/\Lambda)^2}}^{\rm atanh\sqrt{1-(\tau/\Lambda)^2}}d\eta .
\ee
Note that (only) for $\tau=0$ the $\eta$-integral extends from $-\infty$ to $+\infty$. 
We therefore consider instead the action 
\be \label{2}
{\cal S}=\int d\tau \int_{-\infty}^{+\infty}\!{\cal L} \;d\eta  =\int(T-L_2-U) d\tau
\ee
with
\be \label{3}
T(\tau)=\frac{\tau}{2}\int_{-\infty}^{+\infty} (\dt\Phi)^2 d\eta,~~~~~~ 
L_2(\tau)=\frac{1}{2\tau}\int_{-\infty}^{+\infty}(\dz\Phi)^2 d\eta,~~~~~~
U(\tau)=\tau\int_{-\infty}^{+\infty} V(\Phi) d\eta.
\ee
Variation with respect to $\Phi$ then leads to the equation of motion (EOM) $\delta {\cal S}/\delta\Phi=0$:
\be \label{4}
\frac{1}{\tau}\dt \Phi+\dtt\Phi-\frac{1}{\tau^2}\dzz \Phi
+\delta V/\delta\Phi=0.
\ee
Consider the total energy $E(\tau)$ at a given proper time $\tau$
\be \label{6}
E(\tau) \equiv \int((\delta {\cal S}/\delta\dt\Phi)\cdot\dt\Phi -{\cal L})d\eta = (T(\tau)+L_2(\tau)+U(\tau)).
\ee
The total derivative of this energy with respect to proper time $\tau$ is 
$$
\frac{dE}{d\tau}=\frac{\partial E}{\partial\tau}
+\int\left(\frac{\delta E}{\delta\dt\Phi}\dtt\Phi 
+\frac{\delta E}{\delta\Phi}\dt\Phi\right) d\eta 
$$
\be \label{7}
 =\frac{1}{\tau}(T-L_2+U) + 
\tau\int\left(\dtt\Phi-\frac{1}{\tau^2}\dzz\Phi
+\delta V/\delta\Phi\right)\dt\Phi \;d\eta.
\ee
Using EOM (\ref{4}) leads to
\be \label{8}
\frac{dE}{d\tau}=\frac{1}{\tau}(-T-L_2+U)=-\frac{1}{\tau}(E-2U).
\ee
This is solved by
\be
E=\frac{C}{\tau}+\frac{2}{\tau}\int^\tau U d\tau
\ee
with constant $C$.
\subsubsection{Two remarks:}
(A): 
If the potential $V(\Phi)$ averaged over rapidity is independent of proper time $\tau$
(as can be realized for random distributions of $\Phi$
in rapidity space, i.e. during the early phase of an evolution)
such that we have $U\propto\tau$, then a simple solution to (\ref{8}) is obtained 
\be \label{9}
T+L_2= \frac{C_0}{\tau}, ~~~~~~~~~ \mbox{i.e.}~~~ 
E=\frac{C_0}{\tau}+C_1\tau
\ee
with constants $C_0$ and $C_1$.
The first term ( solving the homogenous part of eq.(\ref{8})) 
looks like damping, but it is
a purely geometrical decrease. (Its time integral is still divergent).

(B): If the energy for large proper times $\tau\rightarrow\infty$ approaches a finite constant
$E(\tau)\rightarrow E_\infty$, then eq.(\ref{8}) requires that   
\be  \label{9b}
\mbox{lim}_{\tau\to\infty}\; (T(\tau)+L_2(\tau) - U(\tau)) =0 +\mbox{fluct.} 
\ee
with eventual fluctuations in $U$ and $T+L_2$ (which, however have to add up 
to a smooth total energy $E(\tau)$, because of $\frac{dE}{d\tau}\rightarrow 0$). For proper-time 
averages over at least one period of the fluctuations we have
\be  \label{9c}
\mbox{lim}_{\tau\to\infty}\;< (T(\tau)+L_2(\tau)> =
\mbox{lim}_{\tau\to\infty}\;< U(\tau)> = E_\infty/2.
\ee

(B1): This may be realized such that for large $\tau$ all time derivatives disappear
($T \to 0$) and $L_2 =U \to E_\infty/2$. This represents 
an ensemble of adiabatically shrinking solitons which carry all of the
conserved energy $E_\infty$ (see below). 

(B2): Another way to arrive at an asymptotically constant $E$ appears if 
the gradient terms $L_2$ become negligible after a long time. Then eq.(\ref{9c}) 
requires that asymptotically $<U>$ and $<T>$ are equal and constant in proper time.
This implies that there exist non-vanishing average time-derivatives in the
system which decrease like $1/\sqrt{\tau}$ in proper time.

\subsection{Shrinking solitons at rest in the Bjorken frame}

For a stability condition we omit all time-derivative terms, i.e. put $T=0$
and keep in the EOM only 
\be \label{10}
-\frac{1}{\tau^2}\dzz \Phi+\delta V/\delta\Phi=0.
\ee
This equation describes extended objects whose size in rapidity space
shrinks proportional to increasing proper time, which here simply acts
as a scaling parameter.
The derivative of their total energy $E=L_2+U$ with respect to proper
time is given by
\be \label{11}
\frac{dE}{d\tau}=\frac{1}{\tau}(U-L_2).
\ee
From multiplying EOM (12) with $\dz \Phi$ we have
\be \label{12}
\dz\left(-\frac{1}{2\tau^2}(\dz\Phi)^2+ V(\Phi)\right)=0,
\ee
or, if the integrals are finite,
\be \label{13}
L_2=U,~~~~~~~\mbox{and} ~~~~~~~dE/d\tau=0.
\ee
This result, that for the shrinking solitons $L_2$ and $U$ are independent of proper time, can of course 
be obtained by simply replacing
$\Phi(\eta)$ by $\Phi_\tau(\eta)=\Phi(\tau\eta)$ in the definitions
(\ref{3}). 

However, this only holds in an adiabatic limit, where proper time can be considered as a parameter. 
In fact, for the shrinking solitons
we have $\dt \Phi_\tau =(\eta/\tau) \dz \Phi_\tau$, so
\be \label{13b}
T=\frac{1}{2\tau^2}\int \eta^2 (\dz \Phi(\eta))^2 d\eta.
\ee
So, in the limit of large proper times, $T$ becomes rapidly negligible as compared to the
constants $L_2$ and $U$, and $E$ is asymptotically conserved.

\subsection{ Including a dissipative term}

Genuine dissipation is taken into acount by adding a dissipative term
to the EOM. With dissipation constant $\gamma$, eq.(6) now reads
\be \label{14}
(\gamma+\frac{1}{\tau})\dt \Phi+\dtt\Phi-\frac{1}{\tau^2}\dzz \Phi
+\delta V/\delta\Phi=0.
\ee
Inserting this into eq.(\ref{7}) we have
\be \label{15}
\frac{dE}{d\tau}=-\frac{1}{\tau}(E-2U)-\gamma\tau\int(\dt\Phi)^2d\eta.
\ee
The first term again is purely geometrical, while the rate of energy
loss due to dissipation is given by the second term alone
\be \label{16}
\left(\frac{dE}{d\tau}\right)_{\rm diss}=-\gamma\tau\int(\dt\Phi)^2d\eta.
\ee
It is equal to $-2\gamma$ times the kinetic energy $T$, so we have
\be \label{15a}
\frac{d(T+L_2+U)}{d\tau}=-\frac{1}{\tau}(T+L_2-U)-2\gamma T.
\ee
For the early stages of an evolution, in analogy to (\ref{9}), we expect for the smooth
parts of $T,L_2,$ and $U$
\be \label{9aa}
<T+L_2>= \frac{C_0}{\tau}f(\tau), ~~~<U>=C_1 \tau.
\ee
The function $f(\tau)$ satisfies the differential equation
\be \label{diffeq}
\frac{df}{f}=-2 \gamma \frac{<T>}{<T+L_2>} d \tau.
\ee
The dissipation will exponentially reduce the kinetic gradients as compared to the 
rapidity gradients, so for the smooth parts  $<T>$ and $<L_2>$ we try the ansatz
\be
<T>= \exp(-\alpha\gamma\tau)<L_2>, 
\ee
where we have allowed for an as yet undetermined constant $\alpha$. This leads to
\be
f(\tau)=\left(\frac{1+e^{-\alpha\gamma\tau}}{1+ e^{-\alpha\gamma\tau _0}}\right)^{2/\alpha}
\ee 
if we require that $f(\tau_0)=1$ for all values of $\gamma$, i.e. if for the initial configurations
at $\tau=\tau_0$ the integrals $T+L_2$ are independent of $\gamma$.
This provides an analytical expression for $<L_2>$ 
\be
\label{L2approx}
<L_2>=\frac{C_0}{\tau}\frac{(1+e^{-\alpha\gamma\tau})^{2/\alpha -1}}{(1+ e^{-\alpha\gamma\tau _0})^{2/\alpha}}
\ee
as compared to
explicitly solving the EOM (\ref{14}). We shall see below that its intersection with the linearly rising
potential $<U>=C_1 \tau$ may be used to define a simple estimate for the freeze-out time $\tau_f$.

\subsection{Adiabatic Approximation}
For large values of $\gamma$, such that $\gamma\tau_0\gg 1$, all fluctuating terms are suppressed from
the very beginning, the evolutions get overdamped and we may
approximate the EOM (\ref{14}) by omitting all second-order time derivatives. 
As $\tau >\tau_0$, it is then also sufficient to keep in the dissipative  terms only the constant $\gamma$ 
\be \label{17}
\gamma \;\dt \Phi-\frac{1}{\tau^2}\dzz \Phi
+\delta V/\delta\Phi=0.
\ee
With this equation of motion we now have 
\be \label{18}
\frac{d (L_2+U)}{d\tau}=\frac{1}{\tau}(-L_2+U)-2\gamma T.
\ee
Proceeding as in equations (\ref{9aa}) and (\ref{diffeq}), with $U=C_1 \tau$, we now have
\be \label{L2adia}
L_2= \frac{C_0}{\tau}\exp\left(-2\gamma \int_{\tau_0}^\tau(T/L_2)\; d\tau \right).
\ee
With a suitable ansatz for the ratio $T/L_2$ during the early part of the evolution we may again obtain
a simple expression for $L_2$ and, in connection with $U=C_1 \tau$, for the freeze-out time $\tau_f$. 

\section{ Simple example: $\Phi^4$-model}

The $\Phi^4$-model uses the local potential
\be \label{V}
V(\Phi)=\frac{\lambda}{4} (\Phi^2-1)^2,
\ee
so the action ${\cal S}$ in $(z,t)$-coordinates is
\be \label{20}
{\cal S}=\int
\left(\frac{1}{2}(\partial_t\Phi)^2-\frac{1}{2}(\partial_z\Phi)^2 
-\frac{\lambda}{4} (\Phi^2-1)^2 \right) dzdt.
\ee
For stable static solitons we have \cite{Raja}
\be  \label{s1}
\frac{1}{2}\int (\partial_z\Phi)^2dz=\frac{\lambda}{4} \int(\Phi^2-1)^2
dz=\frac{\sqrt{2\lambda}}{3} \;. 
\ee

\subsection{Shrinking solitons in the Bjorken frame}
After transforming to the Bjorken frame the stability condition (\ref{10}) looks 
like the equation for stable static solutions in the rest frame 
with $\lambda$ replaced by 
$\lambda \tau^2$. So we have according to eq.(\ref{s1})
\be  \label{s2}
\frac{1}{2}\int(\partial_\eta\Phi)^2d\eta=\tau^2
\frac{\lambda}{4} \int (\Phi^2-1)^2 d\eta=\tau\frac{\sqrt{2\lambda}}{3} \;.
\ee
With definitions (\ref{3}), we find 
\be  \label{s3}
L_2=U=E/2= \frac{\sqrt{2\lambda}}{3}= \mbox{const}_\tau.
\ee
So, in the (adiabatic, i.e. $T\rightarrow 0$) limit of large proper times the total energy of the
shrinking solitons in the Bjorken frame approaches a constant which coincides with
their static energy in the $(z,t)$-frame. This represents a realization 
of the above remark (B1) in section II.

It should, however, be
noted that on a lattice these relations are violated if with increasing
proper time the shrinking size 
$(\tau \sqrt{\lambda/2})^{-1}$ of the stable 
solitons approaches the lattice constant (set equal to one). Gradients for single
solitons then cannot exceed the value of 2 (corresponding to a jump in $\Phi$ from -1 to +1 for
neighbouring lattice points). So we have 
\be \label{limit}
L_2 \rightarrow n \frac{2}{\tau}~~~~~~ \mbox{for}~~~~~ \tau\rightarrow\infty,
\ee
where $n$ is the final number of kinks+antikinks
surviving for large proper times. This means that $L_2$ will not approach a constant, but will 
decrease like $1/\tau$, while the kinks' contribution to $U$ rapidly goes 
to zero like $ \tau^{-3}$. 
So, on a lattice, $dE/d\tau =0$ cannot be reached in an 
adiabatic limit for large proper times.
Still, $dE/d\tau \rightarrow 0$ for $\tau\rightarrow\infty$ can be realized 
by balancing fluctuational contributions to 
$T$ and $U$ such that
$E$ is constant, which constitutes a non-adiabatic realization of
the remark (B2) discussed in section II.A.

\subsection{Evolutions on a lattice}
\subsubsection{Symmetric potential}

Let us first consider evolutions in the symmetric potential
\be \label{V+}
V^{(+)}(\Phi)=\frac{\lambda}{4} \left((\Phi^2+1)^2-1 \right),
\ee
which has only one single minimum at $\Phi=0$. (We subtract the 
constant $-1$ to maintain $V(0)=0$). So, a priori, we would not expect any
'phase transition' to occur in this potential.
Initial configurations at $\tau=\tau _0\ll 1$ for the evolutions
consist of values $\Phi_i$ at each lattice point 
$i=0,...,N$, randomly chosen from a normalized Gaussian distribution 
\be \label{Gauss}
f(\Phi)= \frac{1}{\sqrt{2\pi\sigma^2}}\exp(-\frac{\Phi^2}{2\sigma^2})
\ee
with mean square deviation $\bar{\Phi^2}=\sigma^2$.
So, the system looks similar for all
rapidity intervals that make up the $\eta$ lattice.

The integral
\be \label{V1}
\int V^{(+)}(\Phi) d\eta \Longrightarrow N \int V^{(+)}(\Phi) f(\Phi)d\Phi
\ee
for sufficiently small width $\sigma\ll 1$ is approximated by 
\be
N \frac{\lambda}{2} \int\Phi^2 f(\Phi)d\Phi = N \frac{\lambda}{2}\sigma^2.
\ee
So, the initial value of $U^{(+)}$ at $\tau=\tau_0$ is given by 
$N \frac{\lambda}{2}\sigma^2 \tau_0$, while  $U^{(+)}(\tau)$
will be oscillating around the time-averaged $<U^{(+)}>$ given by the straight line
\be \label{U+}
<U^{(+)}> = N \frac{\lambda}{4}\sigma^2 \tau.
\ee
For an estimate of the integral $\int (\dz\Phi)^2 d\eta$ we implement the
first derivatives on the lattice as $\Phi_{i+1} -\Phi_i$
and obtain 
\be \label{deriv}
\frac{1}{2}\int (\dz\Phi)^2 d\eta \Longrightarrow \sum (\Phi_i)^2 -\sum\Phi_{i+1}\Phi_i
=N(\bar{\Phi^2}-\bar\Phi^2).
\ee
For the second equality we have assumed that the two-point correlation function
$\sum\Phi_{i+1}\Phi_i$ factorizes, i.e. that $\Phi$-values at neighbouring lattice points are
uncorrelated. So, the initial value for $L_2$ at $\tau=\tau_0$ is given by
 \be \label{L_2ini}
L_2(\tau_0) = N \frac{\sigma^2}{\tau_0}.
\ee
The initial values for the proper time derivatives $\dt \Phi_i (\tau_0)$ are not relevant
and we can put them to zero. The equations of motion immediately after the onset of the 
evolution build up $T$ such that the smooth sum $T+L_2$ closely follows the line $N\sigma^2/\tau$,
while $T$ and $L_2$ separately oscillate around the time averaged values
 \be \label{L_2}
<L_2>=<T> = N \frac{\sigma^2}{2 \tau}.
\ee
During this early phase of the evolution the potential is unimportant because it is suppressed
by two orders of $\tau$ as compared to $L_2$. 
 The proper time value 
\be \label{tf}
\tau_f^2=2 /\lambda
\ee
marks the intersection of
$<L_2>$ with the potential $<U^{(+)}>$ where $U^{(+)}$ begins to influence the
motion. This marks the end of this first (gradient-dominated) phase which consists
of freely propagating fluctuations.

\begin{figure}[h]
\centering
\includegraphics[width=10cm,height=14cm,angle=-90]{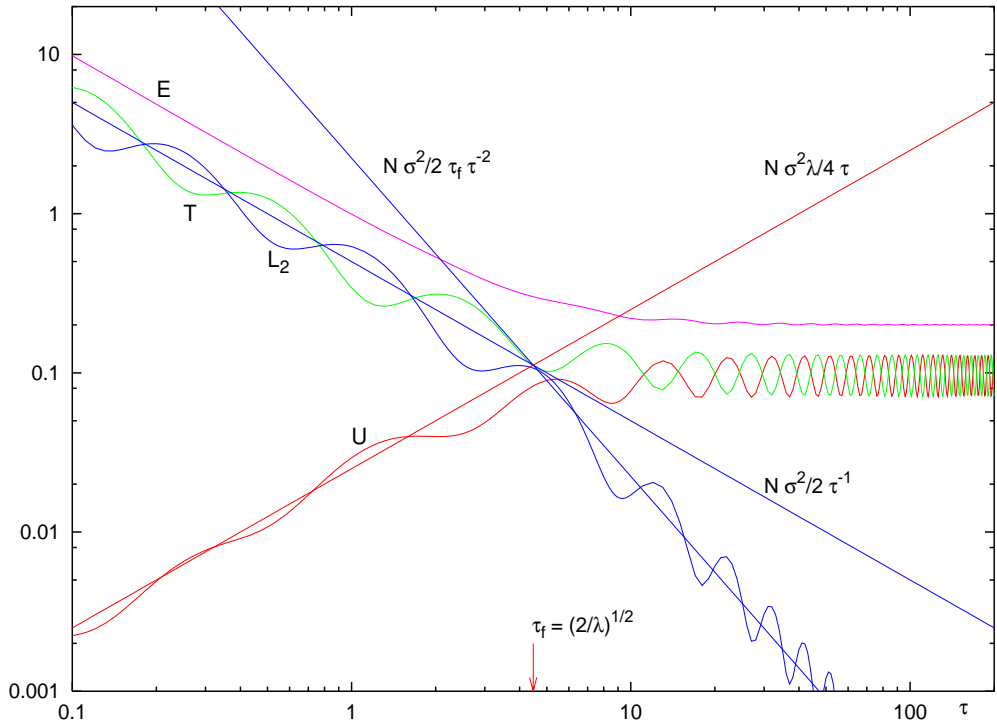}
\caption{Evolution on an $N=100$ lattice of a single event 
according to EOM (\ref{4}) in the $\Phi^4$-model
with the $V^{(+)}$ symmetric potential ($\lambda =0.1$). 
The initial distribution at $\tau=\tau_0$ is Gaussian (\ref{Gauss}) with $\sigma=0.1$.
The transition time $\tau _f$ 
is defined by the intersection of the lines 
$N \sigma^2/(2\tau)$ and $N\sigma^2\lambda\tau/4$.} 
\label{fig1}
\end{figure}

For times $\tau\gg\tau_f$, the gradient terms become irrelevant. 
If, for large $\tau$,
we neglect in the EOM (\ref{4}) the gradient terms altogether, then 
the fluctuations satisfy the simplified EOM 
\be \label{21}
\frac{1}{\tau}\dt \phi+\dtt\phi+\lambda\phi=0
\ee
which is solved by
\be  \label{amplitudes}
\phi\sim \frac{1}{\sqrt{\tau}} \cos \:(\sqrt{\lambda}\tau -\pi/4) + {\cal O}(\tau^{-1}),
\ee
i.e., the amplitudes of the fluctuations decrease like $1/\sqrt{\tau}$.
This implies that both, $U^{(+)}$ and $T$, oscillate with constant amplitude,
and frequency $2\sqrt{\lambda}$, both adding up to an asymptotically 
constant $E_\infty=T+U^{(+)}$,
while $L_2$ rapidly decreases like  $1/\tau^2$.
 
So, for times $\tau\gg\tau_f$, we are left with a system dominated by 
non-propagating fluctuations, or, in other words, with a system of N independently
moving oscillators, all fluctuating with the same frequency $\sqrt{\lambda}$ 
(the 'mass' at $\Phi=0$) and amplitudes decreasing like $1/\sqrt{\tau}$.  
Asymptotically for large proper times the total energy is conserved,
realizing case (B2) discussed above.

Figure \ref{fig1} shows the typical features of an evolution in the $V^{(+)}$-potential
(for $\lambda=0.1$ and $\sigma=0.1$). As long as $\tau_0\ll \tau_f$, 
the actual choice of $\tau_0$ and the initial values of $\dt \Phi_i(\tau=\tau_0)$ 
are unimportant. So, in contrast to our original expectations, the evolution shows
very distinctly a transition from the early gradient-dominated phase to a 
system of $N$ independent oscillators dominated by the local potential.
In the first phase we observe the virial theorem $<T>=<L_2>$, in the second phase
we have $<T>=<U>$. At the transition time $\tau=\tau_f$, which is independent of 
$\sigma$ and $\tau_0$, $<L_2>$ changes continuously from
the $\tau^{-1}$ to $\tau_f\tau^{-2}$ behaviour.

The order of magnitude of $\tau_f$ can be obtained from the following argument:
In the Bjorken frame, the field fluctuations carry an effective wave number
$k/\tau$ and mass $\sqrt{\lambda}$. If $k$ were not limited from above, then for any $\tau$ 
there could be propagating waves with $(k/\tau)^2\geq\lambda$. However, if the system
is characterized by a momentum cut-off, $k<\Lambda$, then for $\tau^2\gg\Lambda^2/\lambda$,
the system decouples into $N$ independent oscillators with the same frequency $\sqrt{\lambda}$.  
On the lattice, $k$ is limited by the finite
lattice constant $a$: $k<\pi/a$. So, (with $a$ put to 1), this provides an estimate
for the freeze-out time $\tau_f$. The statistical average over the initial ensemble
then leads to (\ref{tf}).
From this consideration we conclude that the observed freeze-out is the consequence
of a momentum cut-off, which in the present case is imposed by the lattice. We should, however,
keep in mind that here the lattice constant has a well-defined physical meaning: 
With our choice of the initial ensemble, where values of the field variables $\Phi_i$
at neighbouring lattice sites are uncorrelated, the lattice constant can be identified with
the initial correlation length $\xi$. So, in this case, the freeze-out time should rather be
written as 
\be \label{tfr}  
\tau_f^2=\frac{2}{\lambda\xi^2}.
\ee

\subsubsection{Potential with broken symmetry}
Let us now consider an evolution subject to the potential $V$ (\ref{V}), which is 
characterized by two degenerate minima at $\Phi=\pm 1$ and satisfies $V(\pm 1)=0$.
As before, for sufficiently small width $\sigma\ll 1$ of the initial distribution,
the potential is unimportant during the early phase of an evolution. So,
starting from the same initial distribution as described above, the first part of an 
evolution proceeds
exactly as in the case of the symmetric potential $V^{(+)}$ (\ref{V+}). 
The smooth part of the integral $U^{(-)}=\tau\int V^{(-)}d\eta=\tau\int \left(V(\Phi)-V(0)\right) d\eta$
again is approximated by 
\be \label{U-}
<U^{(-)}>=(-)N \frac{\lambda}{4}\sigma^2 \tau,
\ee
while $<L_2>$ is still given by (\ref{L_2}). 
So, the freeze-out time $\tau_f$ as determined from
\be  \label{freeze}
-<U^{(-)}>=<L_2>
\ee
again is given by (\ref{tf}). 
(Due to the opposite sign of $<U^{(-)}>$ as compared to the case of the symmetrical potential 
$<U^{(+)}>$, this condition here implies that
even for the highest wave numbers $k\sim 1/a$ supported by the lattice, 
the frequencies become imaginary for $\tau>\tau_f$).

For $\tau>\tau_f$,  the gradients which couple the field variables $\Phi_i$ at neighbouring 
lattice sites are suppressed and the onset of the potential drives 
each $\Phi_i$ towards +1 or -1, depending only on the sign of $\Phi_i$ at $\tau\sim\tau_f$.
So, evidently,  $\tau_f$ is also that point in time when the number $n$ of 
kinks+antikinks finally found in the system for $\tau\to\infty$
is determined. Necessarily, because for $\tau>\tau_f$ the $\Phi_i$ at neighbouring sites
no longer interact, the analytical shape of the stable (shrinking) kink is irrelevant, or,
in other words, the size $R=\left(\tau \sqrt{\lambda/2}\right)^{-1}$ of the kinks has 
shrunk below the lattice unit. This provides another interpretation of the freeze-out
time which we could have chosen alternatively to define $\tau_f$. 
Clearly, for $R<1$, kinks and antikinks no longer overlap, so there is no further
annihilation and their total number $n$ is conserved.

\begin{figure}[h]
\centering
\includegraphics[width=10cm,height=14cm,angle=-90]{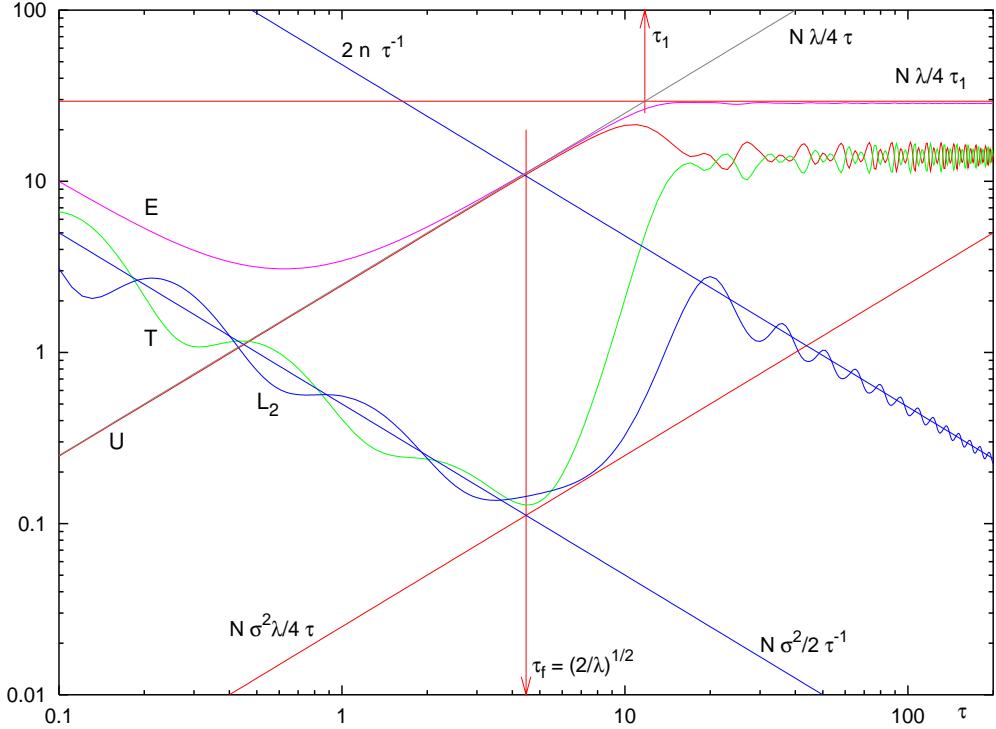}
\caption{Evolution on an $N=100$ lattice of a single event (with $\sigma=0.1$) 
according to EOM (\ref{4}) in the $\Phi^4$-model
with the broken-symmetry potential $V(\Phi)$ (\ref{V})
(for $\lambda =0.1$). 
The freeze-out time $\tau _f=\sqrt{2/\lambda}=4.47$ 
is defined as in figure~\ref{fig1} by the intersection of the lines 
$N \sigma^2/(2\tau)$ and $N\sigma^2\lambda\tau/4$. 
The number $n$ of kinks+antikinks remaining in the final configuration
is tied to the intersection of the lines 
$2n/\tau$ and $N\lambda\tau/4$.
This particular event leads to $n=24$. The arrow  at $\tau=\tau_1$ points to the end-of-roll-down time
as given by eq. (\ref{t1}). Its intersection with the line $N\lambda\tau/4$ marks the finally available
energy $E_\infty$ (in the approximation discussed in subsection III.E). } 
\label{fig2}
\end{figure}

Figure \ref{fig2} shows the typical features of an evolution in the symmetry-broken 
potential $V$ (\ref{V}) (for $\lambda=0.1$), starting from an initial Gaussian 
distribution with $\sigma=0.1$. For proper times larger than the freeze-out time, 
$\tau>\tau_f=\sqrt{2/\lambda}$, $L_2$ switches from oscillating around
$N \sigma^2/(2\tau)$ to oscillating around $2n/\tau$ (with $n=24$ in this particular 
example), while $U$ begins to deviate from $N\lambda\tau/4$ and joins with $T$ 
into combined oscillations around the constant $E_\infty/2$.
These fluctuations consist of the $N$ $\Phi_i$-values oscillating independently
around the vacua $\Phi=\pm 1$ with frequency $\omega_0=\sqrt{2\lambda}$ 
(the 'mass' in the true vacuum) and amplitudes decreasing like $1/\sqrt{\tau}$, 
so $U$ and $T$ asymptotically oscillate with frequency $2\omega_0$ with constant amplitudes.

We can obtain an estimate for the number $n$ by the following consideration: 
As given in (\ref{limit}),
for large proper times $\tau\gg\tau_f$ we expect $L_2$ to approach the limit $<L_2>=2n/\tau$ 
which on the lattice comprises the kinks' contribution to the energy. 
For proper times up to $\tau_f$, $\tau<\tau_f$, the integral $\int V d\eta$ is dominated by the 
constant contribution $V(0)$, i.e. we have $<U>=N\lambda\tau/4$. 
Both constants, $V(0)$ before the transition, and $2n$ after the transition, do not enter into the 
EOM which describes the fluctuations around them and which determines the transition time $\tau_f$.
In normal time $t$, energy conservation would require both constants to be equal. To translate that into the
proper-time frame, we have to extrapolate $<L_2>$ 
backwards to the freeze-out time $\tau_f$ and consider its intersection with $<U>$ at $\tau_f$  
\be \label{nlat}
<U>=<L_2>|_{\tau=\tau_f}~~~~~\mbox{i.e.}~~~~N\frac{\lambda}{4}\tau_f = 2 n   \frac{1}{\tau_f}.
\ee
So we obtain as an estimate for the number $n$ of kinks+antikinks expected to survive,
\be \label{nu}
n=\frac{N}{4}.
\ee
For this consideration the fluctuational contributions
to $U$ and to $L_2$ are unimportant because they are suppressed by order ${\cal O}(\sigma^2)$,
(apart from the fact that they determine $\tau_f$). 
The result (\ref{nu}) is very simple and independent of $\lambda$ (and of $\sigma$ and $\tau_0$,
if chosen sufficiently small). It may be noted that it is just one half of the purely statistical 
Kibble limit $n_K=N/2$.
Numerically, it can be measured as the average kink+antikink number $\bar n$ 
obtained for a large ensemble of events.  
Averaging over 1000 events we find a slight dependence on $\tau_0$ and $\sigma$: for $N=100$,
$\sigma=0.01$ and $\tau_0=0.1, 0.01$ we have $\bar n=25.0 (4.7), 27.6 (5.0)$; 
for $\sigma=0.1$ and $\tau_0=0.1, 0.01$
we find $\bar n =26.1 (4.7), 28.9 (5.1)$, respectively. (Cf. figs.\ref{fig6} and \ref{fig7}). 
(The numbers in brackets give
the corresponding mean deviations).

Although the momentum cut-off implemented through the lattice (which is decisive for the freeze-out time
and instrumental for the occurrence of the transition) can have a genuine physical meaning, it is still
unsatisfactory that the shrinking kinks' contribution to the energy, due to the finite lattice constant,
is given by $2n/\tau$,
and not by their true energy (\ref{s3}) $E= n \cdot 2\frac{\sqrt{2\lambda}}{3}$ which is independent of $\tau$. 
To correct for that lattice artifact
we can replace (\ref{nlat}) by  
\be  \label{ntrue}
N\frac{\lambda}{4}\tau_f =  2n \frac{\sqrt{2\lambda}}{3}
\ee
and, with (\ref{tf}), obtain the result
\be
n=\frac{3}{8} N 
\ee
which again is independent of $\lambda$. We shall see in the following that this is the appropriate
procedure, if the freeze-out occurs before the soliton size has shrunk to the lattice-unit.

\subsection{Temperature and initial distributions}
We have seen from the preceding results that the  
momentum cut-off $\Lambda=\pi/a$, here imposed on the system by the finite lattice constant $a$, is crucial
for the freeze-out time and for the onset of the phase transition. Although, for the initial configurations chosen,
the lattice constant does acquire physical meaning as the initial correlation length, it may seem very
unsatisfactory that the essential physical processes rely on a rather arbitrarily chosen momentum cut-off.
Therefore it is more natural to introduce a cut-off $\Lambda_{\cal T}$ through a temperature $\cal T$ 
characterizing the initial distribution. For this purpose we construct the initial
configurations $\Phi_i$ ($i=1,..,N$) on the lattice, as Fourier transforms 
$$
\Phi_i=\sqrt{2/N}\sum_{k=1}^{N} \sin \left(\frac{\pi}{N} k\cdot i\right) \tilde\Phi_k
$$
of distributions 
$\tilde \Phi_k$ (k=1,..,N)
in momentum space, chosen randomly from a Gaussian deviate $f_G(\tilde\Phi)$ with $k$-dependent width $\sigma_k$,
\be\label{Gausstemp}
f_G(\tilde\Phi_k)=\frac{1}{\sqrt{2\pi \sigma_k^2}} \exp\left(-\frac{\tilde\Phi_k^2}{2\sigma_k^2}\right),
~~~~~~~\mbox{with}~~~~~~~~
\sigma_k^2=\frac{\sigma_0^2}{\sqrt{2\pi \sqrt{\lambda}\cal T}} 
\exp\left(-\frac{(k\pi/(aN))^2}{2\sqrt{\lambda}\cal T}\right).
\ee
In other words, we choose a Boltzmann distribution for the average occupation numbers 
$<n_k>=<\tilde\Phi_k^2> = \sigma_k^2$ for particles with mass $\sqrt{\lambda}$ ( i.e., the 'mass' of the
fluctuations in the symmetric potential $V^{(+)}$ ) and temperature $\cal T$. On a lattice, the upper limit
for $k$ is $N$. So, as long as 
\be \label{LamT}
{\cal T} < \Lambda_{\cal T},~~~~~~\mbox{with}~~~~~~\Lambda_{\cal T}=\frac{\pi^2/a^2}{2\sqrt{\lambda}},
\ee
the lattice cut-off is unimportant.
With $a=1, \lambda=0.1$ we have ${\cal T}< 15.6$. Within that limit we obtain for the width of the 
initial distribution of the field configurations $\Phi_i$ on the 'coordinate' lattice
\be \label{sigma}
\sigma^2=<\Phi^2> =\frac{1}{N}\sum_k <\tilde\Phi_k^2>= \frac{\sigma_0^2}{2\pi},  
\ee
which even for $\sigma_0^2=1$ still is smaller than 1. 
So, our previous arguments which lead to $|<U^{(\pm)}>|= N \sigma^2\lambda/4 \tau$ still hold and we obtain
in place of (\ref{U-})
\be
|<U^{(\pm)}>|=N\lambda \;\frac{\sigma_0^2}{8\pi}\tau.
\ee
For an estimate of $<L_2>$, as in equation (\ref{deriv}) we need the correlation function $<\Phi_{i+1}\Phi_i>$,
which now no longer factorizes, but instead is given by
\be
\frac{1}{N}\sum_{i=1}^{N}<\Phi_{i+1}\Phi_i>=\frac{\sigma_0^2}{2\pi}\exp \left(-\sqrt{\lambda}{\cal T}/2\right).
\ee
So, together with (\ref{sigma}) we obtain for $<L_2>$
\be
<L_2>=\frac{N\sigma_0^2}{4\pi \tau}\left(1-\exp \left(-\sqrt{\lambda}{\cal T}/2\right)\right).
\ee
We again use $|<U^{(\pm)}>|=<L_2>$ to get the freeze-out time
\be \label{tftmp}
\tau_f^2=\frac{2}{\lambda}\left(1-\exp \left(-\sqrt{\lambda}{\cal T}/2\right)\right).
\ee
If we use as an estimate for the average kink+antikink multiplicities $n$ the relation (\ref{nlat}) we find
\be \label{ntlat}
n=\frac{N}{4}\left(1-\exp \left(-\sqrt{\lambda}{\cal T}/2\right)\right).
\ee  
However, for small temperatures $\cal T$, the freeze-out time (\ref{tftmp}) will be much shorter than
the time $\sqrt{2/\lambda}$ when the kinks' size approaches the lattice unit. So it appears more appropriate 
to replace (\ref{nlat}) by (\ref{ntrue}), which accounts for the fact that at times $\tau<\sqrt{2/\lambda}$
the emerging (and shrinking) kinks' energy is constant and given by $2\frac{\sqrt{2\lambda}}{3}$. This leads to
\be \label{nttrue}
n=N\frac{3}{8}\left(1-\exp \left(-\sqrt{\lambda}{\cal T}/2\right)\right)^{1/2}.
\ee 
For small temperatures this results in a square-root law for the multiplicities 
$n/N=\frac{3}{8\sqrt{2}}\sqrt{\sqrt{\lambda}\cal T}$.
\begin{figure}[h]
\centering
\includegraphics[width=8cm,height=14cm,angle=-90]{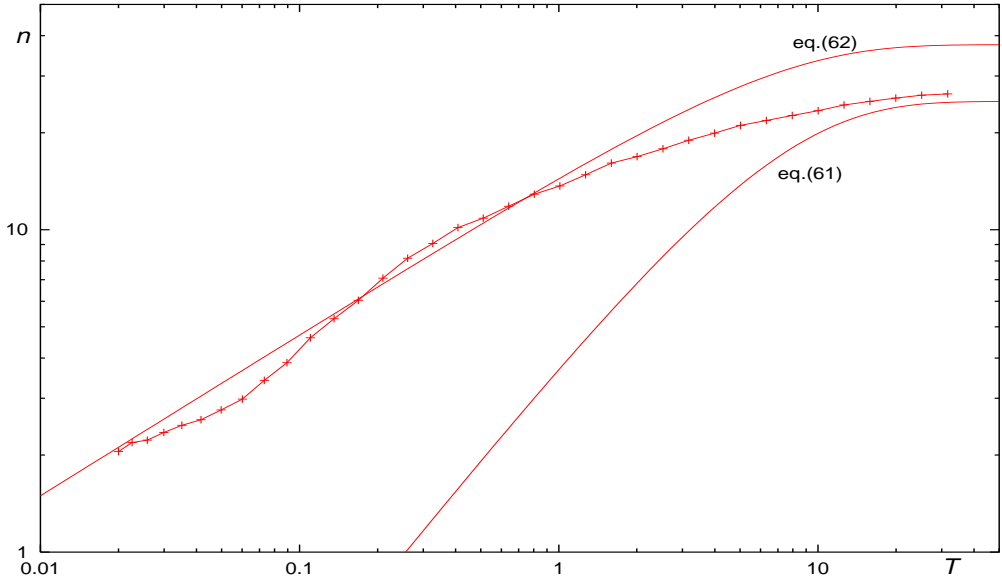}
\caption{Kink+antikink multiplicities $\bar n$ (averaged over 500 events) as function of the initial
temperature $\cal T$ (for $N=100, \lambda=0.1$). The solid lines are calculated from the expressions
(\ref{ntlat}) and (\ref{nttrue}), respectively.}
\label{fig3}
\end{figure}

In figure \ref{fig3} we compare the numerically measured $\bar n$ (averaged over 500 events, $N=100$,
$\lambda=0.1$) as function of initial temperature $\cal T$ with both expressions, (\ref{ntlat}) and (\ref{nttrue}).
Evidently, using the shrinking kink's true constant energy $2\frac{\sqrt{2\lambda}}{3}$, 
provides a good approximation for $\bar n$ in the
low-temperature region, while the use of the lattice expression $2/\tau$ describes only the limit where the
temperature exceeds the cut-off $\Lambda_{\cal T}$ in (\ref{LamT}). These results show that our expressions for the
freeze-out time and the multiplicities $n$ work reasonably well, independently of the physical origin 
of the momentum cut-off. We shall therefore in the following return to the simple case where the momentum-cut-off is 
provided by the lattice constant.

\subsection{Cooling times}
Up to now we have considered the 'sudden quench', i.e. 
the initial 'hot' configuration was exposed from the outset $\tau\geq\tau_0$ 
to a fixed potential.
In general, however, the $\Phi^4$-potential would be of a form like
\be \label{Vtemp}
V(\Phi)=\frac{\lambda}{4} (\Phi^2-v^2(\tau))^2
\ee
with a time- (or temperature-) dependent function $v^2(\tau)$ to drive the phase transition.
Typically, the functional dependence of $v^2$ on proper time $\tau$ would be like 
\be \label{v2}
v^2(\tau)= \tanh (\tau/\tau_c-1),
\ee
parametrized by a cooling time $\tau_c$ which characterizes the time it takes the potential
to approach its 'cold' form.  A possible scenario how the time-scale of this change in the effective
chiral potential may be driven by a confining transition coupled to it has been discussed in the
rapidly expanding Bjorken metric in~\cite{Scavenius}. In the present context, however, we shall 
simply impose a suitable time dependence like (\ref{v2}).

\begin{figure}[h]
\centering
\includegraphics[width=8cm,height=12cm,angle=-90]{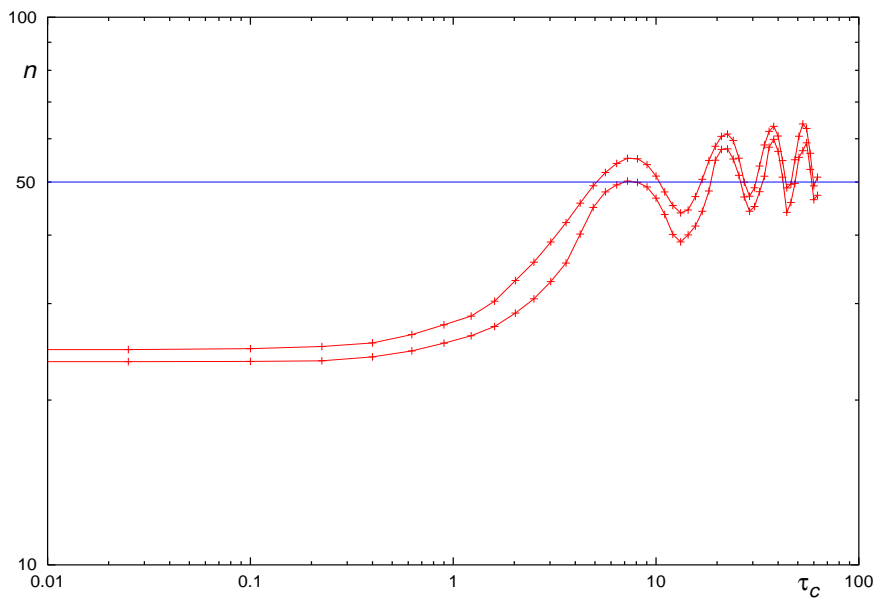}
\caption{Kink+antikink multiplicities $\bar n$ (averaged over 500 events) as function of cooling time
$\tau_c$ (for $N=100, \lambda=0.1$, and for two different choices of the initial width 
$\sigma=0.0001$ and $\sigma=0.01$). The frequency of the oscillations around
$n=N/2$ is $\omega=\sqrt{2\lambda}$.}
\label{fig4}
\end{figure}

We have seen that in the Bjorken frame for a fixed potential the kinetic freeze-out time 
$\tau_f=\sqrt{2/\lambda}$ in (\ref{tf}) 
is obtained independently of the sign of $v^2$, being solely determined by the
coupling constant $\lambda$ and the momentum cut-off which characterizes the system. 
Retracing our arguments which have lead to equations (\ref{U+}) and (\ref{U-}),
now for the $\tau$-dependent potential (\ref{Vtemp}), we obtain
for the freeze-out time
\be \label{tftemp}
\tau_f^2= \frac{2}{\lambda \; v^2(\tau_f)}
\ee
For an ansatz like (\ref{v2}) the solutions $\tau_f^2$ of this equation satisfy $\tau_f^2 > 2/\lambda$.
They approach the lower limit $2/\lambda$ for $\tau_c\rightarrow 0$. For very long cooling times
$\tau_c\rightarrow\infty$, we have $\tau_f/\tau_c\rightarrow 1$. In any case, if $\tau_c $ is smaller
or comparable to $\sqrt{2/\lambda}$, then also the freeze-out time $\tau_f$ is comparable to $\sqrt{2/\lambda}$
and we can apply equation (\ref{nlat}) which determines the kink multiplicities. With the potential
(\ref{Vtemp}) this is now modified as
\be
N\frac{\lambda}{4} v^4 \;\tau_f =  \frac{n}{2}(2v^2)^2 \frac{1}{\tau_f}.
\ee
So, with (\ref{tftemp}),
we obtain the same result $n=N/4$ as before, independently of the value of $v^2(\tau_f)$. 
We conclude, that as long as the cooling time $\tau_c$ is smaller  than the
kinetic freeze-out limit $\sqrt{2/\lambda}$, the resulting multiplicities are almost 
independent of $\tau_c$, and it is  justified to use the sudden quench approximation $\tau_c =0$.

This is no longer true for $\tau_c \gg \sqrt{2/\lambda}$. In that case the potential (\ref{Vtemp}) is
still symmetric with a single ($\Phi=0$) minimum at the time $\tau=\sqrt{2/\lambda}$ when all modes decouple.
Therefore, by the time $\tau=\tau_c$ when the potential develops its degenerate true minima 
with broken symmetry, all field variables $\Phi_i$ roll down into the nearest minimum, independently
of each other. So, for large cooling times, the average number $\bar n$ will approach the Kibble limit $N/2$.
Figure \ref{fig4} shows kink+antikink multiplicities $\bar n$ (averaged over 500 events) 
as function of cooling time
$\tau_c$ (for $\lambda=0.1$, and for two different choices of the initial width 
$\sigma=0.0001$ and $\sigma=0.01$).
The fact that $\bar n$ oscillates with $\omega = \sqrt{2\lambda}$ (the 'cold' mass) shows that the roll down
proceeds only after the potential has essentially reached its final 'cold' form. 

Physically, this is not interesting because the field variables $\Phi_i$ at different lattice sites
do not interact with each other for $\tau \gg \sqrt{2/\lambda}$. We therefore in the following consider
only the sudden quench $\tau_c=0$.

\subsection{Evolution after freeze-out}

At freeze-out the energy present in the rapidity interval is completely converted into kink-antikink pairs.
After freeze-out the roll-down sets in, i.e. the field evolution is dominated by the potential.
Locally, the field picks up kinetic energy $T$ under the influence of the potential $-\frac{\lambda}{2}\Phi^2$
(for $|\Phi|\ll 1$). As long as $|\Phi|$ is sufficiently small the equation of motion is 
\be
\frac{1}{\tau}\dot\Phi+\ddot \Phi -\lambda\Phi =0,
\ee solved by 
\be
\Phi(\tau)=c \;I_0(\sqrt{\lambda}\tau)=c \: \exp(\sqrt{\lambda}\tau)F(\sqrt{\lambda}\tau).
\ee
with $F(x)$ a slowly monotonically decreasing function of $x$.
If $\sigma$ is the (average) length of $\Phi$ at $\tau_f$, then the constant~$c$ is 
$c=\sigma/I_0(\sqrt{\lambda}\tau_f)$.
For the typical roll-down time $\tau_1$ we take
the time when $|\Phi(\tau_1)|=1$ is reached. So, we have
\be \label{t1ex}
\sqrt{\lambda}(\tau_1-\tau_f) = -\ln \sigma + \ln \frac{F(\sqrt{\lambda}\tau_f)}{F(\sqrt{\lambda}\tau_1)}.
\ee
As a first approximation we use
\be \label{t1}
\tau_1-\tau_f = -\frac{1}{\sqrt{\lambda}}\ln \sigma.
\ee 
(This is a lower limit, because the omitted last term in (\ref{t1ex}) is $>0$, and because 
for $|\Phi|\sim 1$ the 4-th order terms in $V$ become important.
But for an estimate, (\ref{t1}) is good enough. The estimate should be best for small $\sigma$.) During this
time interval the spatial volume covered by the Bjorken slab grows linearly with proper time.

The local energy density is dominated by the constant
$\lambda/4$, so by the end of the roll-down the additional energy $(\tau_1-\tau_f)N\lambda/4 $ is available
to be converted into $\sigma$-fluctuations around the true vacua $\Phi=\pm 1$. These fluctuations carry the mass
$\omega_0 = \sqrt{2\lambda}$. So, at the end of the evolution, in addition to the $n$ (anti)kinks formed 
at freeze-out time, we find $\sigma$-mesons emitted from the considered 
rapidity slab with multiplicity
\be \label{nsigma}
n_\sigma=\frac{N\lambda}{4}\frac{(-1)}{\sqrt{\lambda}}\ln \sigma \frac{1}{\sqrt{2\lambda}}
=-\frac{N\ln\sigma}{4\sqrt{2}}.
\ee
Altogether, with $n=N/4$ from (\ref{nu}) we obtain for the multiplicity ratio of 
baryon-antibaryon pairs to $\sigma$-mesons
\be \label{nratio}
\frac{n/2}{n_\sigma}= -\frac{1}{\sqrt{2}\ln\sigma}
\ee    
which is independent of the coupling constant $\lambda$. 

In fig.(\ref{fig2}), in addition to the arrow at freeze-out time $\tau_f=\sqrt{2/\lambda}=4.47$ 
another arrow shows the end-of-roll-down time $\tau_1=\tau_f+\sqrt{10}\ln 10 = 11.75$, as given
in (\ref{t1}) for $\lambda=0.1, \sigma=0.1$. Its intersection with the linearly rising 
$U(\Phi=0)$ defines the energy $N\lambda\tau_1/4$ available at $\tau=\tau_1$. 
It can be seen in fig.(\ref{fig2}) that 
the straight horizontal line through this intersection coincides very closely with the actual numerically
calculated total energy $E$ for times $\tau > \tau_1$.  (The contribution $2n\tau^{-1}$ of the kinks is
negligible (on the lattice) for $\tau>\tau_1$).

The last equation (\ref{nratio}) holds for the initial configurations as described in (\ref{Gauss}). 
If we use  the
initial conditions (\ref{Gausstemp}) which involve a temperature cut-off in momentum space, 
we have instead, with (\ref{nttrue}), (\ref{nsigma}) and (\ref{sigma})
\be
\frac{n/2}{n_\sigma}=-\frac{3\sqrt{2}}{4\ln\sigma}(1-\exp(-\sqrt{\lambda}{\cal T}/2))^{1/2}.
\ee 
For small values of $\sqrt{\lambda}{\cal T}$ this reduces to
\be
\frac{n/2}{n_\sigma}=-\frac{3}{4\ln\sigma}(\sqrt{\lambda}{\cal T})^{1/2}.
\ee

With the potential of the true vacuum at $\Phi=\pm 1$ put to zero, after the roll-down 
there is no further linear rise in the total energy
of the expanding Bjorken slab. Remarkably, the $\tau^{-1/2}$ law (\ref{amplitudes}) of the amplitudes of the
remaining vacuum fluctuations provides an asymptotically constant total energy $E(\tau\rightarrow\infty)=E_\infty$,
 similar to the asymptotically constant (exact) energy stored in the shrinking kinks.

\subsection{Including an explicitly dissipative term}

We have seen that in the Bjorken frame the onset of a phase transition, together with the freeze-out 
of independent kinks, antikinks, and 'mesonic' fluctuations around the true vacua, is basically 
governed by the momentum cut-off which characterizes the system. The 'cooling' of the system 
originates in the decrease of effective wave numbers $k/\tau$ with increasing proper time $\tau$. 
In addition to this kinetic cooling we normally would expect also genuine physical cooling to occur
due to radiative processes which dissipate energy from the excited system into the surrounding vacuum.

\begin{figure}[h]
\centering
\includegraphics[width=10cm,height=14cm,angle=-90]{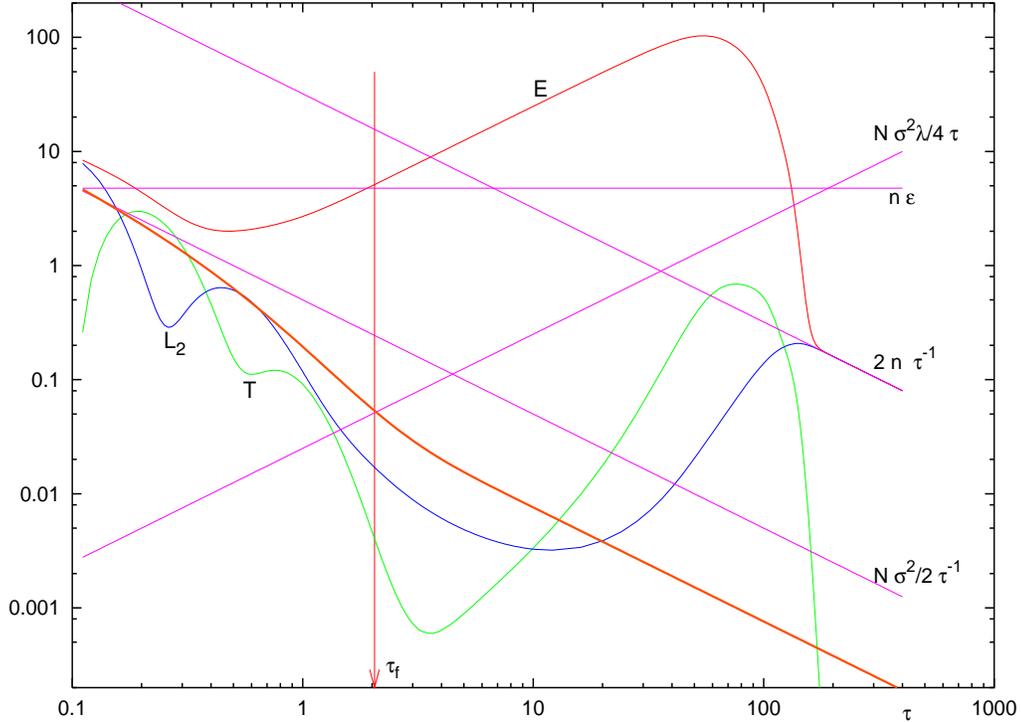}
\caption{Evolution ($N=100,\lambda=0.1,\tau_0=0.1,\sigma=0.1$) according to EOM (\ref{14}) with 
additional dissipation ($\gamma=2$). The thick solid line shows the 
expression (\ref{L2approx})(for $\alpha=1/2$) for the smooth part $<L_2>$. It intersects 
$|<U^{(-)}>| =(N/4)\lambda\sigma^2\tau$ at freeze-out time~$\tau_f=2.05$.
This event produces $n=16$ kinks+antikinks. The constant $16 \:\varepsilon$ 
with $\varepsilon=2\frac{\sqrt{2\lambda}}{3}$ intersects the total energy $E$
near freeze-out time $\tau_f$. The line $16\cdot 2/\tau$ dominates the energy for $\tau > 200$.}
\label{fig5}
\end{figure}

We incorporate this effect by including a finite dissipation constant $\gamma$ into
the EOM (\ref{14}) as described in section II.C.
During the early phase of an evolution, up to the onset of a transition,
this does not affect the linear rise of $<U>$.
On the other hand, the dissipation damps away the rapidity and proper time gradients
such that their smooth sum $T+L_2$ drops below the $N\sigma^2/\tau$ line.
Therefore the freeze-out time $\tau_f$ is reduced as compared to the case $\gamma=0$.
Correspondingly, the average number $\bar n$ of kinks+antikinks produced is also
reduced. As an example, figure~\ref{fig5} shows the evolution of one
event for $\gamma=2$ (with $\sigma=0.1, \tau_0=0.1$). This event leads to $n=16$ kinks+antikinks. 
Applying the approximation (\ref{L2approx}) (with $\alpha=1/2$) 
for an average $<L_2>$, (which in figure~\ref{fig5} is shown by
the solid line), we find as in (\ref{freeze})
from its intersection with the linearly rising $|<U^{(-)}>| =(N/4)\lambda\sigma^2\tau$ the freeze-out time
$\tau_f=2.05$ (as compared to $\tau_f=\sqrt{2/\lambda}=4.47$ for $\gamma=0$). At this early time the
emerging kinks have not yet shrunk to lattice-unit size, so we use their true constant energy 
$\varepsilon=2\frac{\sqrt{2\lambda}}{3}$
to extract from the total energy $E$ at time $\tau_f$ their multiplicity $n=E(\tau_f)/\varepsilon$.
In figure~\ref{fig5} this is indicated by the horizontal line at $16 \:\varepsilon$, which may be seen to intersect
$E$ (which near $\tau_f$ is dominated by the linear rise of potential $<U>$) near proper time $\tau_f$.
The roll-down phase which follows for $\tau>\tau_f$ is characterized by a peak in the proper 
time derivatives $T$, followed by a rapid decrease in $U$, as the configuration settles in the true vacua. 
Finally the total energy is dominated by 
the (lattice-artifact kink energies) $n\cdot 2/\tau$ behaviour of $L_2$ for $n=16$. 
During and after the roll-down phase all fluctuations have been damped away by the dissipative term.

\begin{figure}[h]
\centering
\includegraphics[width=8cm,height=12cm,angle=-90]{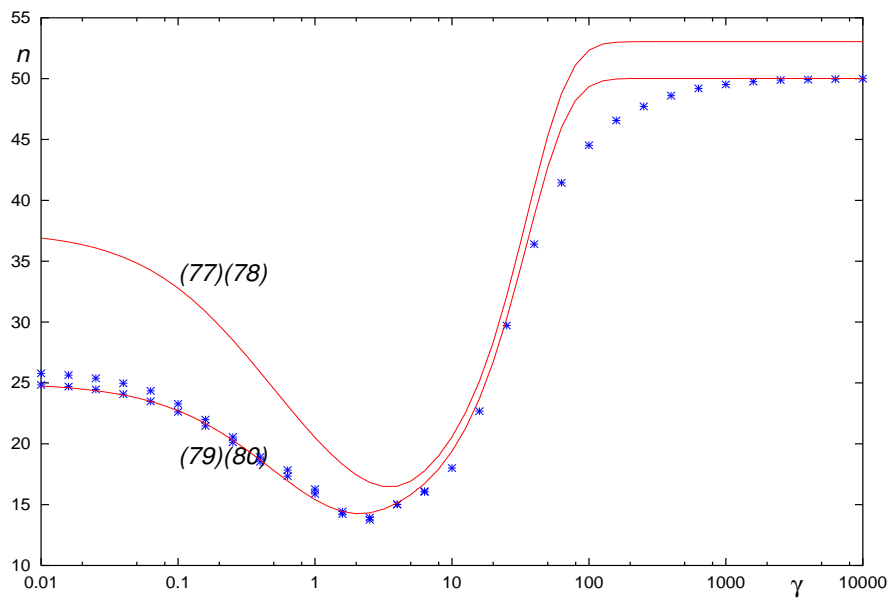}
\caption{Numerically measured kink+antikink multiplicity $\bar n$ (averaged over 1000 random events) as 
functions of damping constant $\gamma$
for initial time $\tau_0=0.1$ (and two values for the width of the initial distribution
$\sigma=0.1$ and $\sigma= 0.01$) .
The solid lines show the results (for $\alpha=1/2$) of the analytical 
approximations eqs. (\ref{ff}),(\ref{nftrue}) and
(\ref{nflat}),(\ref{L2var}), respectively. }
\label{fig6}
\end{figure}

This pattern changes for very large values of the dissipation constant $\gamma$. 
As $\gamma\tau_0$ approaches unity, the 
proper time derivatives $T$ get strongly suppressed already from the beginning of the evolution
such that $T$ stays small and $L_2$ dominates the total energy. For $\gamma\tau_0 \gg 1$,
$L_2$ is determined by its initial value alone, it approaches the line 
\be \label{L_2f}
L_2= N \frac{\sigma^2}{ \tau},
\ee
i.e. it takes twice the value (\ref{L_2}) it has for $\gamma=0$. Correspondingly, the average number $\bar n$ 
of kinks+antikinks approaches the Kibble limit  ${\bar n}\rightarrow N/2$. In other words, the
field configuration is basically frozen from the very beginning at $\tau=\tau_0$, the evolution
consists only of the field variables $\Phi_i$ at each lattice point drifting to the nearest true minimum
at $\Phi=\pm 1$. So, in that limit, the notion of a freeze-out time becomes meaningless.

Still it is interesting to employ the analytical expression (\ref{L2approx}) for $<L_2>$ to determine $\tau_f$ 
from 
\be
\label{ff}
\frac{N\sigma^2}{\tau_f}\frac{(1+e^{-\alpha\gamma\tau_f)^{2/\alpha-1}}}
{(1+ e^{-\alpha\gamma\tau _0})^{2/\alpha}} = N \sigma^2 \frac{\lambda}{4}\tau_f,
\ee 
and the number $n$ of kinks+antikinks from (\ref{ntrue})
\be \label{nftrue}
n=\frac{3}{8}N \sqrt{\frac{\lambda}{2}}\:\tau_f.
\ee
This provides the limits 
$n\rightarrow 3N/8$ for $\gamma\rightarrow 0$,  and $n\rightarrow 3N\sqrt{2}/8$ for $\gamma\tau_0 \gg 1$
and a minimum of $n$ near $\gamma\tau_0\sim 0.5$.  
Figure~\ref{fig6} shows the resulting $n$ 
as function of the dissipation constant $\gamma$, for $\alpha=1/2$, in comparison
with the numerically measured $\bar n$ (averaged over 1000 events, for $\lambda=0.1$, $\tau_0=0.1$
and two values for the width of the initial distribution
$\sigma=0.1$ and $\sigma= 0.01$). Although the general features of $\bar n$ are nicely reproduced, the limits 
for $\gamma\rightarrow 0$  and  $\gamma\tau_0 \gg 1$ are not correct. This is no surprise because in both limits
the true kink energy $\varepsilon$ should be replaced by the lattice result, i.e. 
(\ref{nftrue}) should be replaced by (\ref{nlat})
\be \label{nflat}
n=N\frac{\lambda}{8}\tau_f^2.
\ee
This then leads to the correct 
(lattice) limits 
$n\rightarrow N/4$ for $\gamma\rightarrow 0$, and $n\rightarrow N/2$ for $\gamma\tau_0 \gg 1$. 
It is interesting to note that a quite satisfactory parametrization of the results can be obtained
by using (\ref{nflat}) for all
values of $\gamma$, but replacing $<L_2>$ on the left-hand side of (\ref{ff}) by
\be \label{L2var}
\frac{1+e^{-\alpha\gamma\tau_f}}
{\tau_f \;(1+ e^{-\alpha\gamma\tau _0})^2}= \frac{\lambda}{4}\tau_f.
\ee
This comparison is also shown in fig.\ref{fig6}, again for $\alpha=1/2$.
Evidently, it provides remarkably good agreement for $\gamma$-values up to $\gamma\tau_0<5$.

\subsection{Adiabatic approximation}
\begin{figure}[h]
\centering
\includegraphics[width=8cm,height=13cm,angle=-90]{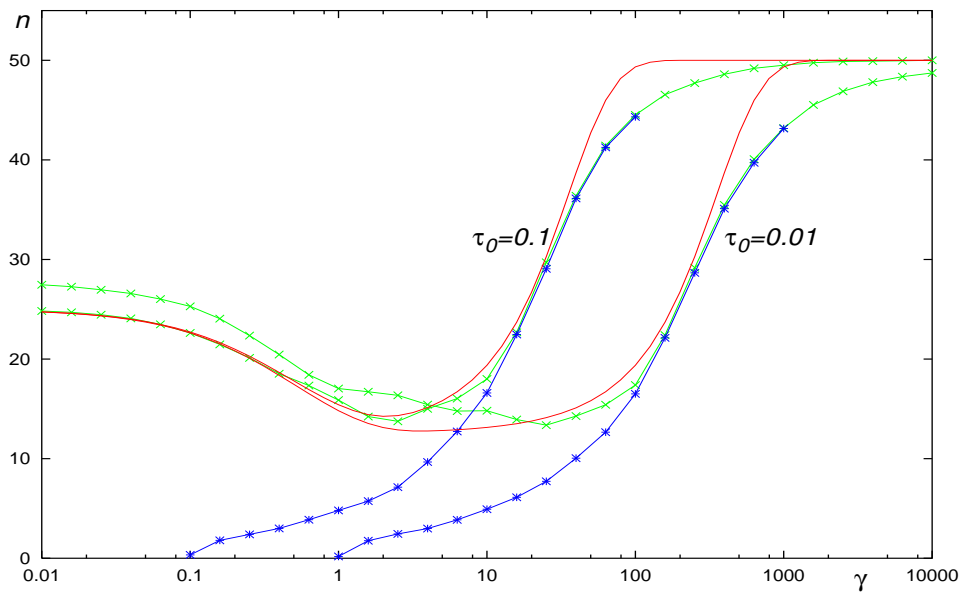}
\caption{Numerically measured kink+antikink multiplicities $\bar n$ 
(averaged over 1000 random events) as 
functions of damping constant $\gamma$
for two different initial times $\tau_0=0.1,0.01$ ( and initial width 
$\sigma=0.01$ ) as obtained from the adiabatic EOM (\ref{17}). For comparison, the non-adiabatic
results are also included.
The solid lines show the result of the analytical approximation (\ref{nflat}),(\ref{L2var})
(with $\alpha=1/2$) for the same two cases.}
\label{fig7}
\end{figure}

For large values of the dissipation constant $\gamma$ it should be sufficient to employ the adiabatic 
approximation as described in section II.D. 
Evolving the initial configurations with the purely dissipative EOM (\ref{17}), leads to the 
kink+antikink multiplicities $\bar n$ (averaged over 1000 events, for $N=100$, and $\lambda=0.1$)
shown in figure \ref{fig7} for two initial times $\tau_0=0.1, 0.01$
(and initial Gaussian width $\sigma=0.01$). For comparison, the non-adiabatic results for the same
two cases are included.

It is apparent that the adiabatic approximation works very well for $\gamma\tau_0 >1$.
It can also be observed that for $\gamma\tau_0< 10^{-2}$ the adiabatic evolutions produce no
kinks at all, i.e. the configurations become so smooth at a very early time (when the 'size' of the
kinks is still very large), such that the complete configuration rolls down into the same true minimum.
The solid lines show the result of the analytical approximation (\ref{nflat}),(\ref{L2var}) for both cases
$\tau_0=0.1, 0.01$.

\section{Conclusion}
The dynamics of interacting fields in rapidly expanding systems is characterized by
effective wave numbers decreasing with time. This is conveniently modeled by transforming
from space-time $(x,t)$ to the rapidity - proper-time  $(\eta,\tau)$ (Bjorken) frame. 
In the equations of motion for the evolving fields,
the rapidity gradients then explicitly carry the inverse proper-time factors, and 
an additional first-order proper-time derivative term appears which resembles dissipative
dynamics, although no genuine physical dissipation is present. In the $(\eta,\tau)$ 
frame, the system still is characterized by a lagrangian which, however, explicitly depends 
on proper time. So, the energy is not conserved, but may decrease or rise during the evolution 
of a field configuration. This is especially interesting for those cases where the system is able 
to undergo a 'phase transition', where 
initial configurations through a cooling process settle down into the true vacua. 
Usually, cooling or heating is
provided by energy exchange with external degrees of freedom (heat bath), 
but in the  $(\eta,\tau)$ frame the transition is solely driven by proper time. 

Systems characterized by multiple different degenerate vacua are of specific interest
because during the course of the phase transition it must be decided how the 
different vacua get populated. Field configurations connecting different vacua will finally
show up as defects, kinks, or solitons, and their multiplicities can be 
physically significant and observable.

The mechanism of how the transition is effectuated by proper time in the Bjorken frame
is quite transparent. The term which comprises the (rapidity-)gradients contribution $L_2$ to the total
energy carries a factor $1/\tau$, while the potential-energy part $U$ carries a factor $\tau$.
Therefore, the evolutions at very early times will be dominated by $L_2$ independently of the potential,
while, if the gradients in the field configurations are limited by a momentum-cut-off~$\Lambda$,
then for very late times the gradients will become unimportant, propagation in rapidity space will end,
and the field variables evolve locally according to the local potential. Evidently, there will exist a freeze-out
time $\tau_f$ which separates both regimes. It is conveniently defined as the time when $L_2$ equals $U$
(where suitable care has to be taken to subtract eventual constants in the definition of the interaction).
Clearly, the momentum cut-off ( or the finiteness of $L_2$ ) is a necessary condition for the transition
to occur, and the actual value of the freeze-out time will depend decisively on $\Lambda$.  
Therefore it is important that the momentum cut-off reflects a genuine physical characteristic of the 
initial field ensemble, typically an initial temperature or inverse correlation length.
This then, however, implies that the freeze-out time can be calculated very simply from the initial
ensemble. 

Stable solitons in the Bjorken frame shrink with increasing proper time, or, in other words,
their stable size relative to the expanding system decreases. But remarkably, their total energy is
constant and equal to their static total energy in the usual (z,t) frame.
This allows to estimate their multiplicities from the total energy available at freeze-out
time (including eventual constants in the definition of the interaction). This means that we may
obtain estimates for the average soliton(+antisoliton) multiplicities  directly from the form 
of the initial ensemble and the static lagrangian which determines their mass. 

A convenient way to introduce a momentum cut-off is to implement the system on a lattice. Then, however,
we have to make sure that the lattice cut-off does not interfere with the increasing gradients which
characterize the shrinking solitons at $\tau_f$. Alternatively, we have to use the lattice expression for the 
soliton's energy if it has shrunk to lattice-unit size.

We have investigated these features in the (1+1)-dimensional $\Phi^4$-model. We find that the simple
analytical expressions for $\tau_f$ and $n$ agree very nicely with numerically measured freeze-out times 
and kink+antikink multiplicities. We have also used Boltzmann ensembles with temperatures well below and
up to the lattice cut-off. Employing the appropriate correlation function we again find good agreement
with the numerical averages. For low temperatures we obtain a square-root law $n\propto \sqrt{\cal T}$.
For temperatures approaching the lattice cut-off the influence of the appropriate soliton energy 
becomes evident. We show that within the physically interesting region it is sufficient to consider
the sudden quench where the initial ensemble is exposed from the beginning to the fixed cold potential.
After the freeze-out the additional energy picked up by the expanding rapidity slice is converted into
mesonic fluctuations. A simple estimate of the roll-down time provides an expression for their average
multiplicities.

In addition to the 'free' evolutions in the Bjorken frame we have also investigated the inclusion
of additional genuine dissipation with variable damping constant $\gamma$. It is very satisfactory 
to find that all the previous considerations can be applied to this case, that the decrease in 
the freeze-out times caused by the damping can be represented in simple analytical expressions
and that the multiplicities $n$ again reflect the soliton energies at freeze-out time. Finally we
find that the validity of the adiabatic approximation, (where all second order time-derivatives are
dropped), is limited by $\gamma\tau_0>1$ where $\tau_0$ denotes the proper time at which the initial ensemble
is prepared and the evolution starts.

It will be most interesting to look at the corresponding features in the (3+1)-dimensional
O(4)-model considered previously as a model for baryon-antibaryon production in relativistic
heavy-ion collisions and compare the results with recent experimentally observed nucleon-antinucleon
multiplicities and with statistical models. The scheme may also be useful for applications 
in cosmology.       
\nopagebreak
\acknowledgments{The author appreciates many helpful discussions with Jorgen Randrup and Hendrik Geyer.
He is especially indebted to Hans Walliser for a careful reading of the manuscript. It is a pleasure to 
thank the Nuclear Theory Group at Lawrence Berkeley Laboratory, and the Physics Department at 
Stellenbosch University for the warm hospitality extended to him.  }

\end{document}